\documentclass[intlimits,twoside,a4paper]{article}

\usepackage{amsmath,amssymb}
\usepackage{graphicx}
\usepackage{wrapfig}
\usepackage{siunitx}
\usepackage{cite}

\usepackage[T2A]{fontenc}
\usepackage[cp1251]{inputenc}

\usepackage[eqsecnum]{cmpj2}

\issue{2012}{15}{3}{33601}

\doinumber{10.5488/CMP.15.33601}

\title[Magneto-sensitive elastomers]{Effects of particle distribution on mechanical properties of
magneto-sensitive elastomers \\ in a homogeneous magnetic field}

\author[D.~Ivaneyko\textsl{ et al}.]{D. Ivaneyko\refaddr{label1,label2},
                            V. Toshchevikov\refaddr{label2,label3},
                            M. Saphiannikova\refaddr{label2},
                            G. Heinrich\refaddr{label1,label2}}
\addresses{
\addr{label1} Institute of Materials Science, Technical University
of Dresden, 7 Helmholtz Str., 01069 Dresden, Germany
\addr{label2}
Leibniz Institute of Polymer Research Dresden, 6 Hohe Str., 01069
Dresden, Germany
\addr{label3} Institute of Macromolecular
Compounds, Russian Academy of Science, Bolshoi Prospect 31, V.O.,
Saint-Petersburg, 199004, Russia
}

\date{Received November 18, 2011, in final form March 26, 2012}

\authorcopyright{D. Ivaneyko, V. Toshchevikov, M. Saphiannikova, G. Heinrich, 2012}

\begin{document}

\maketitle

\begin{abstract}
We propose a theory which describes the mechanical behaviour of
magneto-sensitive elastomers (MSEs) under a uniform external
magnetic field. We focus on the MSEs with isotropic spatial
distribution of magnetic particles. A mechanical model is used in
which magnetic particles are arranged on the sites of three
regular lattices: simple cubic, body-centered cubic and hexagonal
close-packed lattices. By this we extend our previous approach
[Ivaneyko~D. et al., Macromolecular Theory and Simulations, 2011,
\textbf{20}, 411] which used only a simple cubic lattice for
describing the spatial distribution of the particles. The
magneto-induced deformation and the Young's modulus of MSEs are
calculated as functions of the strength of the external magnetic
field. We show that the magneto-mechanical behaviour of MSEs is
very sensitive to the spatial distribution of the magnetic
particles. MSEs can demonstrate either uniaxial expansion or
contraction along the magnetic field and the Young's modulus can
be an increasing or decreasing function of the strength of the
magnetic field depending on the spatial distribution of the
magnetic particles.

\pacs 61.41.+e, 67.30.er, 61.43.Bn
\end{abstract}

\section{Introduction}

Magneto-sensitive elastomers (MSEs), also known as
magnetorheological elastomers, are high-tech materials that can
change their shape and mechanical behaviour under external
magnetic fields~\cite{Filipcsei07}. Nowadays, MSEs have found a
wide range of industrial applications in controllable membranes,
rapid-response interfaces designed to optimize mechanical systems
and in automobile applications such as adaptive tuned vibration
absorbers, stiffness tunable mounts and automobile suspensions~\cite{Carlson00,Ruddy07,Deng07,Chen08b,Ni09,Dong09,Xu10}.

Usually, MSEs consist of micron-sized magnetic particles dispersed
within an elastomeric matrix. The spatial distribution of the
particles inside an elastomer can be either isotropic or
anisotropic (chain-like, plane-like), depending on the method of
preparation~\cite{Zhang08}. MSEs with isotropic distribution of
magnetic particles are synthesized by cross-linking of a polymer
melt with well-dispersed magnetic particles without any external
field. To obtain an MSE with chain-like distributions of
particles, one should apply a strong uniaxial external magnetic
field to a polymer melt before and during its cross-linking~\cite{Filipcsei07}. Using a complex magnetic field with a
rotating vector of magnetic strength or a strong shear flow
before the cross-linking procedure, one can produce an MSE with
plane-like distributions of particles~\cite{Kulichikhin09}.

The magnetostriction effect (i.e., magneto-induced deformation) and
the change of mechanical moduli under external magnetic field
are the most significant properties of the MSEs~\cite{Guan08}. The
deformation of the MSEs can be either positive (elongation) or
negative (contraction) with respect to the direction of the applied
external magnetic field. Magnetostriction and the mechanical
moduli of the MSEs have been investigated by experimental~\cite{Bednarek99,Coquelle05,Martin06,Abramchuk07,Stepanov07,Guan08},
theoretical~\cite{Raikher00,Borcea01,Kankanala04,Diguet10} and
simulation~\cite{Coquelle06,Stepanov08,Raikher08,Raikher09}
studies. In theoretical studies of the mechanical behaviour of the
MSEs, different analytical approaches were proposed, which can be
divided into two groups: continuum-mechanics approach and
microscopic approach. In the continuum-mechanics approach,
electromagnetic equations are coupled with appropriate
mechanical deformation equations. Thus, macroscopic homogeneity of
magnetic media is assumed. This approach predicts positive
magnetostriction for such complex magnetic object as MSEs.
However, the continuum-mechanics approach is not capable of describing
a local discrete spatial distribution of particles.

In the case of microscopic approach, magnetic particles are
considered to be separated by a non-magnetic matrix. Dipole-dipole
interaction between the particles leads to pair-wise attraction
and to repulsion of the magnetic particles depending on their mutual
positions. Since dipole-dipole interactions are very sensitive to
the particle positions, spatial distribution of the particles
inside the matrix strongly affects the sign of magnetostriction, as it was shown in the simulation~\cite{Raikher09} and in the experiment~\cite{Safronov11}. For
instance, MSEs with isotropic spatial distribution of particles
demonstrate an expansion along the magnetic field, while MSEs with
chain-like distribution of particles demonstrate a uniaxial
contraction~\cite{Filipcsei07}.

Recently, we have studied magnetostriction of the MSEs with
isotropic and anisotropic spatial distributions of magnetic
particles  within the framework of a microscopic approach~\cite{Ivaneyko11}. To describe spatial distribution of particles
we have used a regular rectangular lattice model, which permits to
consider ``isotropic'', chain-like and plane-like structures of
particles. Such a regular rectangular lattice model  predicts a negative magnetostriction of
MSE for all distributions of particles. Our prediction for negative magnetostriction is in agreement
with experimental works~\cite{Jolly96a,Zhou04,Coquelle05} for
chain-like distribution. However, predictions of the rectangular
lattice, which degenerates into cubic lattice for isotropic
spatial distribution of particles, contradict the experimental
data for magneto-induced deformation of the MSEs with an
isotropic distribution of magnetic particles~\cite{Zhou04,Diguet10}. The regular rectangular lattice
is probably not capable of sufficiently describing the isotropic distribution.
The reason is that any regular lattice is intrinsically
anisotropic, since one always finds the lattice directions with
different regular distances between the lattice sites. However, in
spite of this feature, some regular lattices can possibly be used
to reasonably well model the mechanical behaviour of the isotropic
MSE. Therefore, in the present study we consider different
lattices to describe isotropic distributions of magnetic particles
in an MSE: simple cubic, body-centered cubic and hexagonal
close-packed lattices. For these three different lattice models we
examine magnetostriction and Young's modulus of the MSE in the
presence of an external magnetic field. We construct the free
energy which consists of elastic and magnetic energies. The
Neo-Hooke law is used to describe entropic non-linear elasticity
of polymer chains.

\section{Free energy of MSE}

\subsection{General equations}

In this section we derive an expression for the free energy of an
MSE. Analysis of the free energy will permit us to study the
mechanical properties of the MSE. Typical MSEs are characterized
by a strong elastic coupling between the particles and the matrix.
This coupling is due to the fact that magnetic particles are much
larger than the mesh size of the polymer network. Since our paper
is devoted to the static mechanical properties of MSEs, we
consider only average displacements of particles coupled to the
matrix. Under these assumptions the free energy of the deformed
MSE under external magnetic field can be written in the form:
\begin{equation} \label{F1}
F=F_\mathrm{el}+F_\mathrm{m}\,.
\end{equation}

The first part $F_\mathrm{el}$ is the elastic energy of a deformed MSE
due to the entropic elasticity of polymer chains. In the present
work we extend a formalism presented in reference~\cite{Ivaneyko11} for linear deformations of MSEs to non-linear
deformations. To calculate the elastic part of the free energy
$F_\mathrm{el}$ of an MSE under finite deformation as a function of the
strain we use the approach of a continuous medium. This approach
means that a sample is divided into representative volumes
of an elastic matrix which contain a large number of particles.
Under this assumption, the free energy $F_\mathrm{el}$ of an MSE under
non-linear deformations can be expressed through the Neo-Hooke law
that can be written in a general form~\cite{Szabo02}:
\begin{equation}
F_\mathrm{el}=\frac{G_0}{2}(I_1-3),
\end{equation}
where the material parameter $G_0$ is the effective shear modulus
of an MSE, $I_1$ is the first scalar invariant of the Finger
strain tensor $\mathbf{B}$: $I_1=\mathrm{tr}\, \mathbf{B}$. The value of
$G_0$ includes contributions of different possible effects into
the elastic energy appearing under elongation of a sample:
reinforcement of an elastic matrix by hard particles due to
non-uniform deformation of the matrix between the particles
(strain amplification), possible adhesion of a polymer matrix onto
the surfaces of hard particles, etc. However, we do not discuss here
how the value of $G_0$ depends on these effects, since this task
is a special problem in the theory of elasticity for isotropic
reinforced rubbers~\cite{Vilgis09}. Instead we use $G_0$ as a
phenomenological parameter of the theory assuming that it can be
extracted from experimental data for elasticity of an MSE in the
absence of a magnetic field. In our theory we assume an
incompressible elastomeric matrix with Poisson's coefficient
$\nu=1/2$. Due to the axial symmetry with respect to the external
magnetic field $\vec{\textbf{H}}$, the MSE will provide an
uniaxial elongation along $\vec{\textbf{H}}$. In this case, the
first scalar invariant $I_1$ has the form:
\begin{equation}
I_1=\sum_i\lambda_i^2\,,
\end{equation}
where $\lambda_x$, $\lambda_y$, $\lambda_z$ are the elongation ratios
for the deformation of an elastomer in the three principal
directions.

The mechanical response of an elastomer to the magnetic field is
characterized by the value of the strain $\varepsilon = \Delta
l/l$, where $\Delta l$ and $l$ are the elongation and the original
size, respectively, of an elastomer along the direction of the
magnetic field ($x$-axis). The condition of constant volume for
elastomers~\cite{Treloar58,Doi86} permits us to relate the
elongation ratios $\lambda_x$, $\lambda_y$, $\lambda_z$ as follows:
\begin{eqnarray}
\lambda_x=1+\varepsilon, \qquad
\lambda_y=\lambda_z=1/\sqrt{1+\varepsilon}\,.\label{RA1}
\end{eqnarray}

Within the framework of Neo-Hooke approximation, the elastic energy for
elastomer as a function of $\varepsilon$ can be written as:
\begin{equation}
F_\mathrm{el}=\frac{E_0}{6}\left[(1+\varepsilon)^2+\frac{2}{1+\varepsilon}-3\right]
=\frac{E_0\varepsilon^2}{2}+O(\varepsilon^3),\label{FEL}
\end{equation}
where $E_0$ is the Young's modulus of a filled matrix: $E_0=3G_0$~\cite{Varga06}. In the present theory, the value of the Young's
modulus $E_0$ of the filled elastomer will be considered as a
parameter in calculations.

\begin{figure}[!t]
\centerline{
\includegraphics[width=6cm]{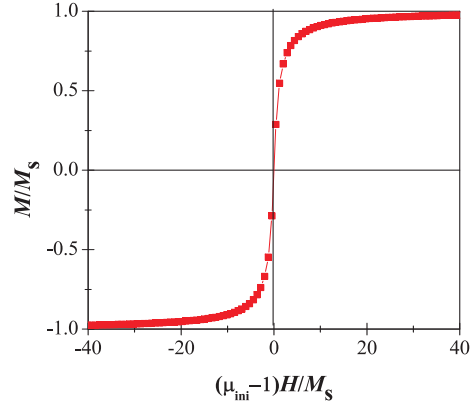}
} \caption{Reduced
magnetization $M/M_\mathrm{s}$ as a function of a reduced magnetic
field $h=(\mu_\mathrm{ini}-1)H/M_\mathrm{s}$. Magnitude of a reduced
magnetization increases with an increasing magnetic field and tends
to saturation magnetization $M_\mathrm{s}$, when $H\rightarrow
\infty$.}\label{fig1}
\end{figure}

The second part of the free energy in equation (\ref{F1}) $F_\mathrm{m}$
arises from the potential energy of magnetic particles placed in
an external magnetic field. Application of a magnetic field
induces an average magnetic moment in each particle. The values of
the induced magnetic moments in the magnetic particles depend on
the material of the particles. Magnetic particles that are usually
used in preparing MSEs have micron-sizes and a multi-domain
magnetic structure. Nevertheless, they are found to show very
narrow hysteresis cycles which indicates a soft magnetic
behaviour. In this case, the dependence of magnitude of the
particle magnetization $M$ on the field strength $H$ can be
described in a good approximation by the Fr\"{o}hlich-Kennely
equation~\cite{Jiles98,Bossis99}
\begin{equation}
M=\frac{M_\mathrm{s}(\mu_\mathrm{ini}-1)H}{M_\mathrm{s} + (\mu_\mathrm{ini}-1)|H|}\,,
\label{FK}
\end{equation}
where $M_\mathrm{s}$ is the saturation magnetization and $\mu_\mathrm{ini}$
is the magnetic permeability of the particles. The values of the
saturation magnetization and magnetic permeability for carbonyl
iron particles are well established in experiment and are equal
to $M_\mathrm{s}\approx$~\SI{1582}{\kilo\ampere/\meter} and $\mu_\mathrm{ini}
\approx 21.5$ with the average diameter of particles of
470$\pm$\SI{180}{\nano\meter}~\cite{Arias06}. Similar values were
obtained for particles of the size of \SI{2}{\micro\meter}:
$M_\mathrm{s}$=\SI{1990}{\kilo\ampere/\meter} and $\mu_\mathrm{ini} = 132$
\cite{Bossis99}. Equation (\ref{FK}) can be rewritten in terms
of dimensionless parameters: reduced magnetic field
$h=(\mu_\mathrm{ini}-1)H/M_\mathrm{s}$ and reduced magnetization $M/M_\mathrm{s}$ in the following form:
\begin{equation}
\frac{M}{M_\mathrm{s}}=\frac{h}{1 + |h|}\,. \label{FK1}
\end{equation}
The dependence of $M/M_\mathrm{s}$ on $h$ is given in
figure~\ref{fig1}. Equation (\ref{FK1}) is used below for
calculations of mechanical characteristics of MSEs in an external
magnetic field.

The interaction energy of magnetic particles in an external
magnetic field consists of two contributions: the dipole-dipole
interaction energy between the particles and the dipole-field
interaction energy~\cite{Jackson98,Froltsov03}. However, under our
assumptions given by equation (\ref{FK}), the dipole-field
interaction energy $-(\vec{\textbf{M}}\cdot\vec{\textbf{H}})$ is
independent of the strain $\varepsilon$, since the bulk
magnetization of the sample $\vec{\textbf{M}}$ is assumed to be
independent of strain $\varepsilon$. Therefore, the dipole-field
interaction energy does not provide a contribution to the
mechanical characteristics and it can be excluded from our
considerations. Thus, in our
approach the magnetic energy $F_\mathrm{m}$ includes  only dipole-dipole interaction energy per unit volume,
which can be written as:
\begin{equation}
F_\mathrm{m}=-\frac{1}{V}\frac{\mu_\mathrm{r}\mu_0}{ 4\pi}\sum_{ij}
\left[\frac{3(\vec{\textbf{m}}_i\cdot\vec{\textbf{R}}_{ij})(\vec{\textbf{m}}_j\cdot\vec{\textbf{R}}_{ij})}{|\vec{\textbf{R}}_{ij}|^5}
-\frac{(\vec{\textbf{m}}_i\cdot\vec{\textbf{m}}_j)}{|\vec{\textbf{R}}_{ij}|^3}\right],\label{FM}
\end{equation}
where $\mu_0$ is permeability of the vacuum, $V$ is the volume
of the sample, $\mu_\mathrm{r}$ is the relative permeability of the medium.
In the present work we consider an elastomeric matrix to be
non-magnetic. Therefore, everywhere below we set $\mu_\mathrm{r}=1$. Here
$\vec{\textbf{m}}_i$ and $\vec{\textbf{m}}_j$ are dipole moments
of  $i$-th and $j$-th magnetic particles, $\vec{\textbf{R}}_{ij}$
is the radius vector that joins the $i$-th and $j$-th  magnetic
particles. The radius vectors $\vec{\textbf{R}}_{ij}$ depend on
the macroscopic deformation $\varepsilon$. To relate
$\vec{\textbf{R}}_{ij}$ with $\varepsilon$ we use an approximation
of affinity of deformation~\cite{Treloar58,Doi86}, as well as
incompressibility of elastomers:
\begin{eqnarray}
(R_{ij})_x&=&(R_{ij}^0)_x\lambda_x=(R_{ij}^0)_x(1+\varepsilon),\label{RA}\\
(R_{ij})_y&=&(R_{ij}^0)_y\lambda_y=(R_{ij}^0)_y(1+\varepsilon)^{-\frac{1}{2}},\label{RB}\\
(R_{ij})_z&=&(R_{ij}^0)_z\lambda_z=(R_{ij}^0)_z(1+\varepsilon)^{-\frac{1}{2}},
\label{RC}
\end{eqnarray}
where $(R_{ij})_{\xi}$ and $(R_{ij}^0)_{\xi}$ are the components
of vectors that join two magnetic particles after and before
deformation, respectively, $(\xi=x,y,z)$. Summation in equation~(\ref{FM}) is performed over all pairs of particles. To simplify
the calculations, we consider regular spatial distributions of
particles. In particular, we use three different lattices in order
to study the effects of particle distribution on the
magneto-mechanical properties of the MSEs in a homogeneous
magnetic field.

\subsection{``Isotropic'' distributions of magnetic particles: different lattice models}

\begin{figure}[!t]
\centerline{
\includegraphics[width=13cm]{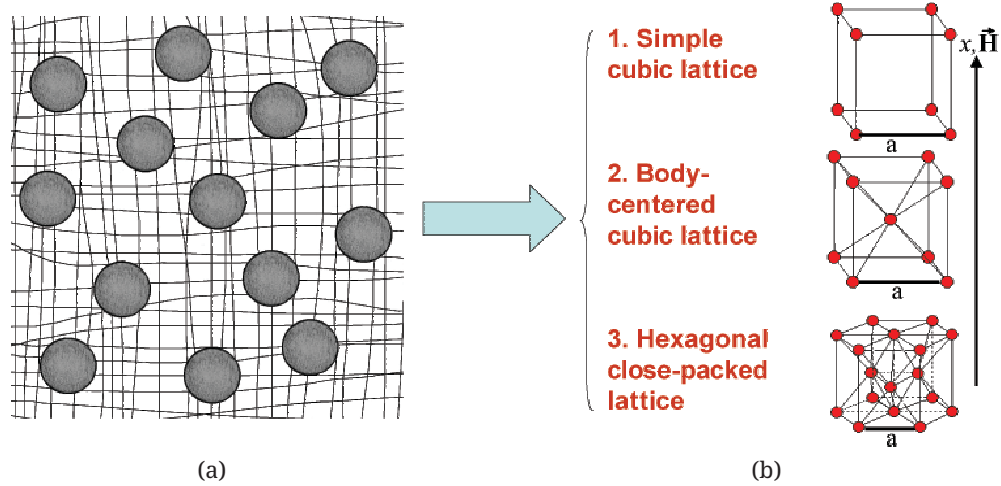}
}  
$\phantom{ffffffffffffffff}$(a)$\phantom{ffffffffffffffffffffffffffffffff}$(b)\\
\caption{Schematic illustration of an MSE with isotropic spatial
distribution of particles (a) and the lattice models (simple
cubic, body-centered cubic and hexagonal lattices) for modelling
the mechanical properties of an MSE (b).}\label{fig2}
\end{figure}

In contrast to our previous study~\cite{Ivaneyko11}, which used
only a cubic lattice to describe the isotropic distribution of
magnetic particles inside the MSE, we consider here different
lattice models (see figure~\ref{fig2}). In these models, it is
assumed that the magnetic particles in an isotropic non-deformed
MSE are located at the sites of simple cubic (SC), body-centered
cubic (BCC) and hexagonal close-packed (HCP) lattices. Let $a$ be the edge length in the three lattices [figure~\ref{fig2}~(b)].

For simplicity we assume that all particles are the same and have
a spherical form; $r$ is the radius of particles. The value of $r$
characterizes the average size of particles in a real elastomer.
Then, the volume fraction, $\phi$, of the particles distributed in
the matrix is given by:
\begin{equation}\label{A1}
\phi = \gamma\frac{\upsilon_0}{a^3}\,,
\end{equation}
where $\upsilon_0=\frac{4}{3}\pi r^3$ is the volume of a particle.
The factor $\gamma$ depends on the type of the lattice:
\begin{eqnarray}
\gamma_{\rm SC} = 1,\qquad \gamma_{\rm BCC} =
2,\qquad \gamma_{\rm HCP} = 2\sqrt{2}.
\end{eqnarray}

Due to the symmetries of the infinite lattices to be used
(translational symmetry and the axial symmetry around the vector
$\vec{\textbf{H}}$), the induced magnetic dipoles
$\vec{\textbf{m}}_i$ and $\vec{\textbf{m}}_j$ are directed along
the external field $\vec{\textbf{H}}$ [$x$-axis in figure~\ref{fig2}~(b)] and their absolute values are identical $m_i=m_j=
\upsilon_0M$, with $M$ being the magnetization of each particle.
Besides, a regular arrangement of the magnetic particles on the
sites of the lattices makes it possible to simplify the summation in the equation~(\ref{FM}). To perform summation over indexes $i$ and $j$ we
choose $j$-th particle and make summation over index $i$. Due to
the translational symmetry for infinite lattices, the contribution
to the magnetic energy of the dipole-dipole interaction of a given
$j$-th particle with other particles around it does not depend on
the number $j$. Thus, the double sum over pairs of indexes $i$ and
$j$ is reduced to a simple sum over index $i$ multiplied by $N$,
where $N$ is the number of all particles. Then, equation~(\ref{FM})
can be rewritten in the form:
\begin{equation}
F_\mathrm{m}= - u_0 \upsilon_0^2c\left(\frac{M}{M_\mathrm{s}}\right)^2\sum_{i}
\left[\frac{3(\vec{\textbf{R}}_{i})_x^2-|\vec{\textbf{R}}_{i}|^2}{|\vec{\textbf{R}}_{i}|^5}\right],\label{FM1}
\end{equation}
where $c=N/V$ is the concentration of particles. Here, for
convenience we calculate the radius vectors from the origin of a
lattice and simplify the notation as
$\vec{\textbf{R}}_{ij}\equiv\vec{\textbf{R}}_{i}$. We introduce
the parameter $u_0$:
\begin{equation}
u_0=\frac{\mu_0 M_\mathrm{s}^2}{ 4\pi}\,,
\end{equation}
that defines the characteristic density of energy of magnetic
interaction. For $M_\mathrm{s} \approx
2\times10^6$~\SI{}{\ampere/\meter} we obtain $u_0=
4\times10^{5}~\SI{}{\pascal}$. Below, we will show that mechanical
behaviour of an MSE in the magnetic field is determined by a
dimensionless parameter $E_0/u_0$, i.e., by the ratio between
characteristic values of the elastic and magnetic energies.

\begin{figure}[!ht]
\centerline{
\includegraphics[width=0.8\textwidth]{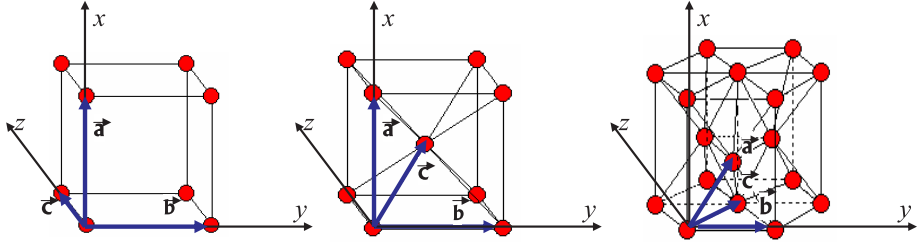}
}
\caption{(Color online) Basis vectors $\vec{a}$, $\vec{b}$ and $\vec{c}$
introduced for the simple cubic, body-centered cubic and hexagonal
close-packed lattices with the unit edge length,
$|\vec{b}|=1$.}\label{fig3}
\end{figure}

The radius vector $\vec{\textbf{R}}_{i}$ in equation (\ref{FM1})
is related with the radius vector $\vec{\textbf{R}}_{i}^0$ in a
non-deformed MSE by equations (\ref{RA})--(\ref{RC}). For simple
cubic, body-centered cubic and hexagonal close-packed lattices, the
value $\vec{\textbf{R}}_{i}^0$ can be presented in the form:
\begin{equation}
\vec{\textbf{R}}_{i}^0=a\cdot\vec{\textbf{r}}_{i}\,. \label{RR}
\end{equation}
The dimensionless vector $\vec{\textbf{r}}_{i}$ runs now over all
sites of the infinite lattice with the edge length $a=1$ except
$\vec{\textbf{r}}_{i}=0$. Using the Fr\"{o}hlich-Kennely
equation~(\ref{FK}) for reduced magnetization $M/M_\mathrm{s}$, we
express the magnetic energy $F_\mathrm{m}$ as a function of a reduced
magnetic field $h$ and the strain $\varepsilon$:
\begin{equation}
F_\mathrm{m}= u_0 \phi^2\left(\frac{h}{1+|h|}\right)^2
f(\varepsilon),\label{UJ3}
\end{equation}
where the dimensionless function $f(\varepsilon)$ has the
following form:
\begin{equation}
f(\varepsilon)=-\frac{1}{\gamma}(1+\varepsilon)^{\frac{3}{2}}\sum_{r_i\neq0}
\frac{2(1+\varepsilon)^3(r_{i})_x^2-(r_{i})_y^2-(r_{i})_z^2}
{\left[(1+\varepsilon)^3(r_{i})_x^2+(r_{i})_y^2+(r_{i})_z^2\right]^{\frac{5}{2}}}\,.\label{FFF}
\end{equation}
We recall that the vector $\vec{\textbf{r}}_{i}$ runs over the sites
of different lattices with $a=1$. The value of function
$f(\varepsilon)$ depends on the type of the lattice, since the
components of the vectors $\vec{\textbf{r}}_{i}$ depend on the
lattice. The vector $\vec{\textbf{r}}_{i}$ for three lattices can
be expressed in the following form:
\begin{equation}
\vec{\textbf{r}}_{i} = \vec{a}\cdot i_a+\vec{b}\cdot
i_b+\vec{c}\cdot i_c\,,
\end{equation}
where $i_a$, $i_b$ and $i_c$ are
the integer numbers, $\vec{a}$, $\vec{b}$ and $\vec{c}$ are the
basis vectors. The basis vectors are presented in figure~\ref{fig3}
and have their coordinates:
\begin{itemize}
  \item Simple cubic lattice
\begin{eqnarray}
\vec{a} = (1,0,0),\hspace{0.5cm}\vec{b} = (0,1,0),\hspace{0.5cm}
\vec{c} = (0,0,1);
\end{eqnarray}
  \item Body-centered cubic lattice
  \begin{eqnarray}
\vec{a} = (1,0,0),\hspace{0.5cm} \vec{b} = (0,1,0),\hspace{0.5cm}
\vec{c} =\left(\frac{1}{2},\frac{1}{2},\frac{1}{2}\right);
\end{eqnarray}
  \item Hexagonal close-packed lattice
  \begin{eqnarray}
\vec{a} =
\left(\sqrt{\frac{2}{3}},\frac{1}{2},\frac{1}{2\sqrt{3}}\right),\hspace{0.5cm}
\vec{b} = (0,1,0), \hspace{0.5cm} \vec{c}
=\left(0,\frac{1}{2},\frac{\sqrt{3}}{2}\right).
\end{eqnarray}
\end{itemize}

For different lattices, the function $f(\varepsilon)$ was analyzed
numerically. Below we consider the magnetostriction and the
Young's modulus of the MSE in the presence of an external magnetic
field depending on the type of the lattice model.

\subsection{Free energy as a function of deformation}

\begin{figure}[!ht]
\centerline{\includegraphics[height=4.1cm]{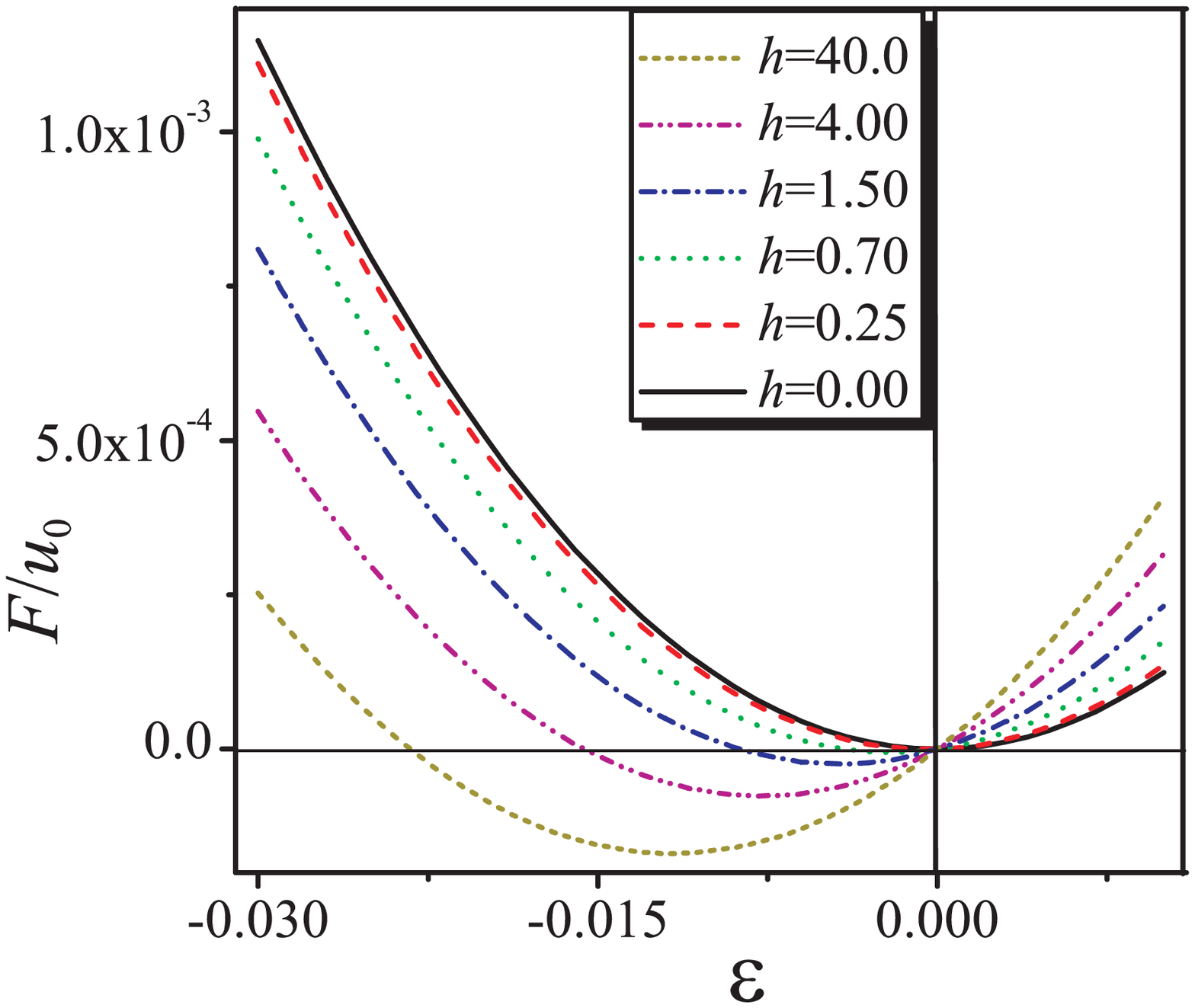}
\includegraphics[height=4.1cm]{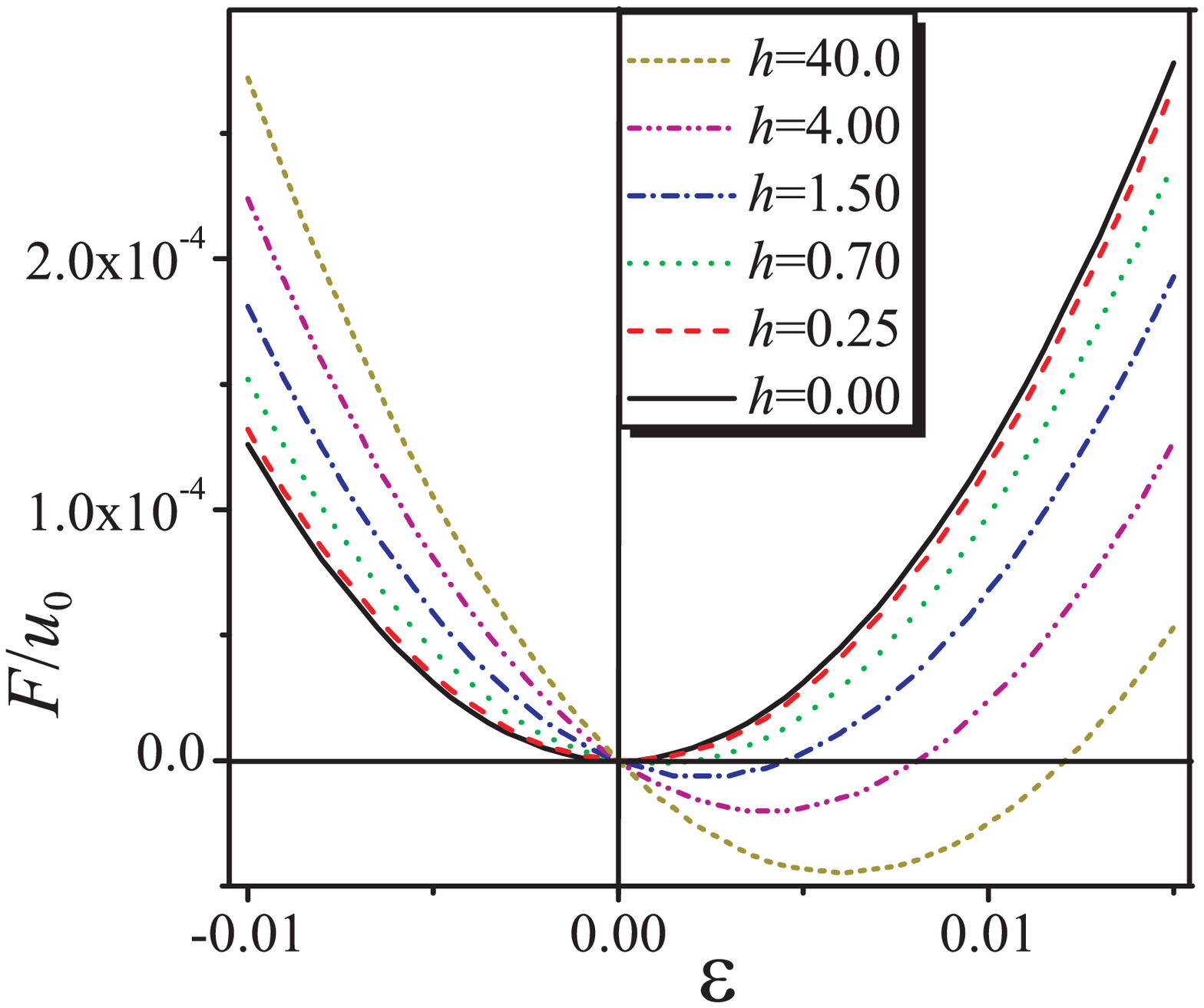}
\includegraphics[height=4.1cm]{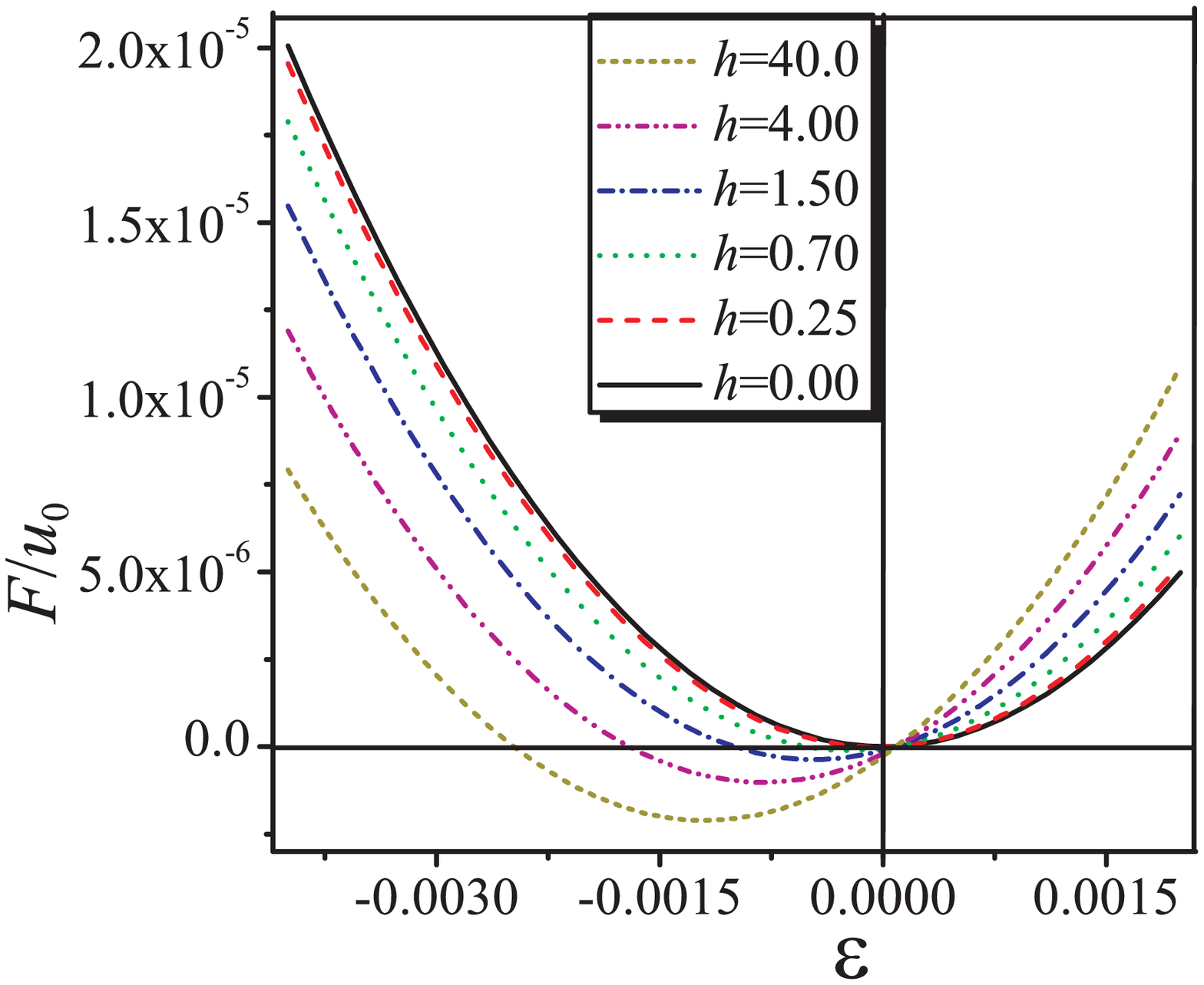}}
\hspace{0.17\textwidth} (a) \hspace{0.31\textwidth} (b) \hspace{0.3\textwidth} (c)
\caption{(Color online) Reduced free energy, $F/u_0$, of an isotropic MSE as a
function of the strain $\varepsilon$ calculated for three types of
initial lattices: simple cubic lattice (a), body-centered cubic
lattice (b) and hexagonal  lattice (c) at different values of the
reduced magnetic field $h=(\mu_\mathrm{ini}-1)H/M_\mathrm{s}$.}\label{fig4}
\end{figure}

Using equation (\ref{FEL}) for  elastic  energy and equation~(\ref{UJ3}) for magnetic energy, we obtain free energy of the MSE in the following form:
\begin{equation}
F=\frac{E_0}{6}\left[(1+\varepsilon)^2+\frac{2}{1+\varepsilon}-3\right]+u_0
\phi^2\left(\frac{h}{1+|h|}\right)^2 f(\varepsilon),\label{FEN}
\end{equation}
where the values of the function $f(\varepsilon)$ vary for the
simple cubic, body-centered cubic and hexagonal close-packed
lattices. We calculate the reduced free energy, $F/u_0$, of an
isotropic MSE as a function of the strain $\varepsilon$ for the
three types of initial lattices at fixed values $\phi=0.05$ and
$E_0/u_0=2.5$ and at different values of the reduced magnetic
field $h=0, 0.25, 0.70, 1.5, 4, 40$ that correspond to the values
$M/M_\mathrm{s}=0, 0.2, 0.4, 0.6, 0.8, 1.0$, respectively. One can
see in figure~\ref{fig4} that application of the magnetic field
leads to the shift of the minimum of  free energy. The minimum
of free energy determines the equilibrium elongation of an
MSE, $\varepsilon_{\rm eq}$. One can see that the value and the
sign of the $\varepsilon_{\rm eq}$ depend on the lattice:
$\varepsilon_{\rm eq} <0$ for the simple cubic and hexagonal
close-packed lattices and $\varepsilon_{\rm eq} > 0$ for the
body-centered cubic lattice. This means that both simple cubic and
hexagonal close-packed lattices predict a contraction of an
isotropic MSE along the magnetic field $\vec{\textbf{H}}$
($\varepsilon_{\rm eq} <0$), whereas the body-centered cubic
lattice predicts an expansion of an isotropic MSE along the
magnetic field $\vec{\textbf{H}}$ ($\varepsilon_{\rm eq} >0$).

The analysis of $\varepsilon_{\rm eq}$ as a function of the
reduced magnetic field $h$, the volume fraction $\phi$ and
parameter $E_0/u_0$ for different lattices are given below.

\section{Mechanical behaviour of MSEs in a homogeneous magnetic field}

\subsection{Magnetostriction of the MSE}

\begin{figure}[!t]
\centerline{
\includegraphics[height=4.1cm]{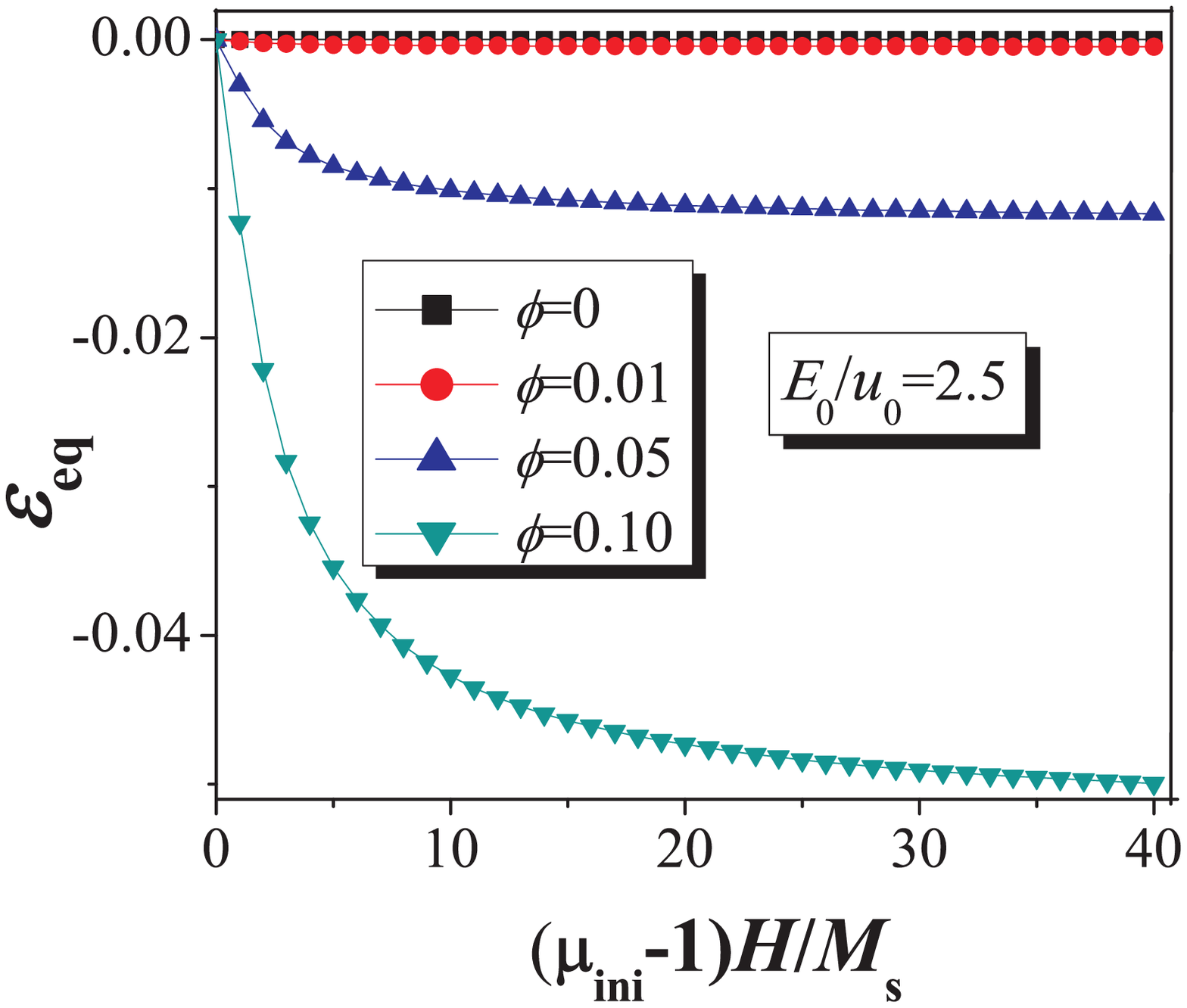}
\includegraphics[height=4.1cm]{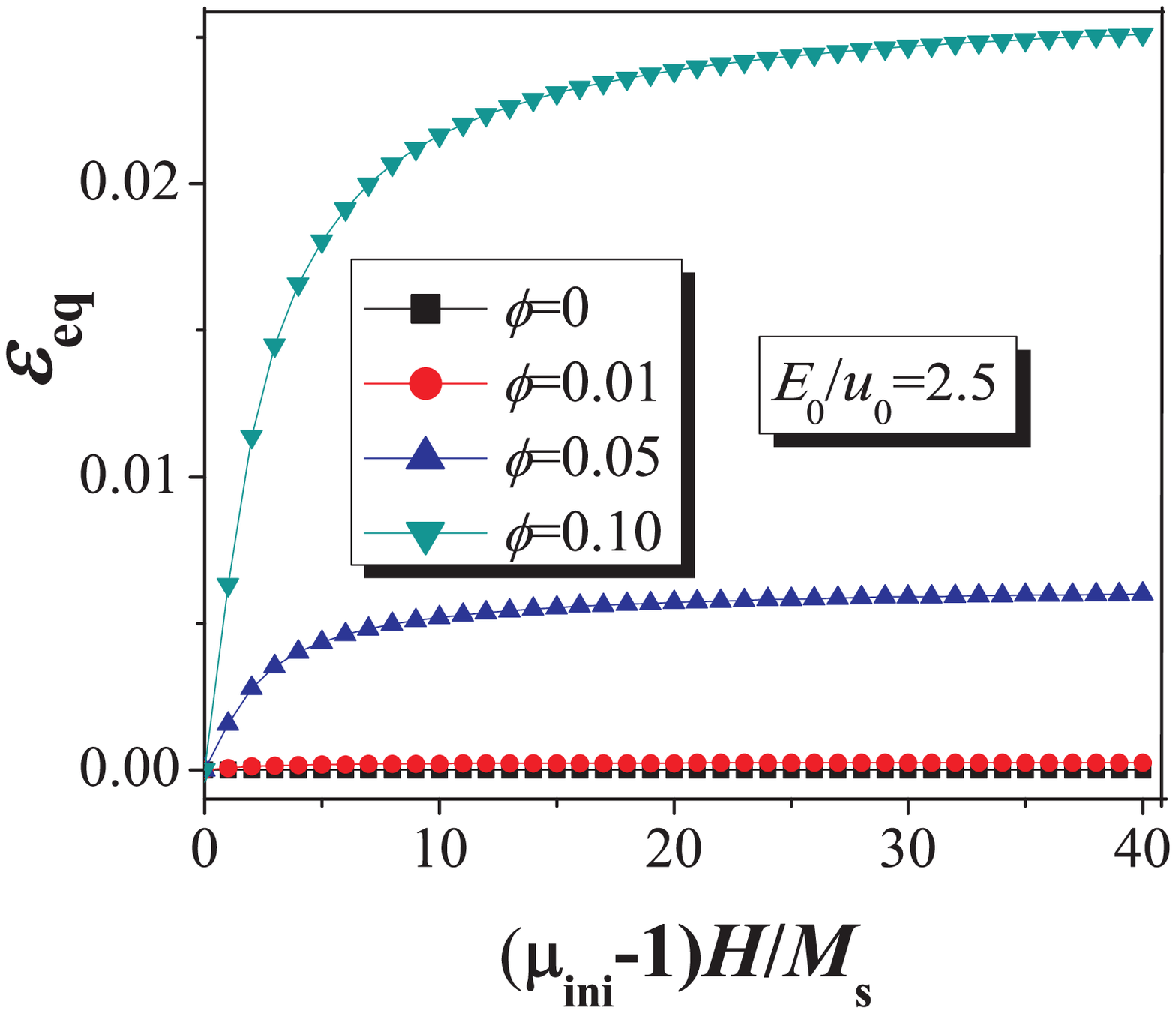}
\includegraphics[height=4.1cm]{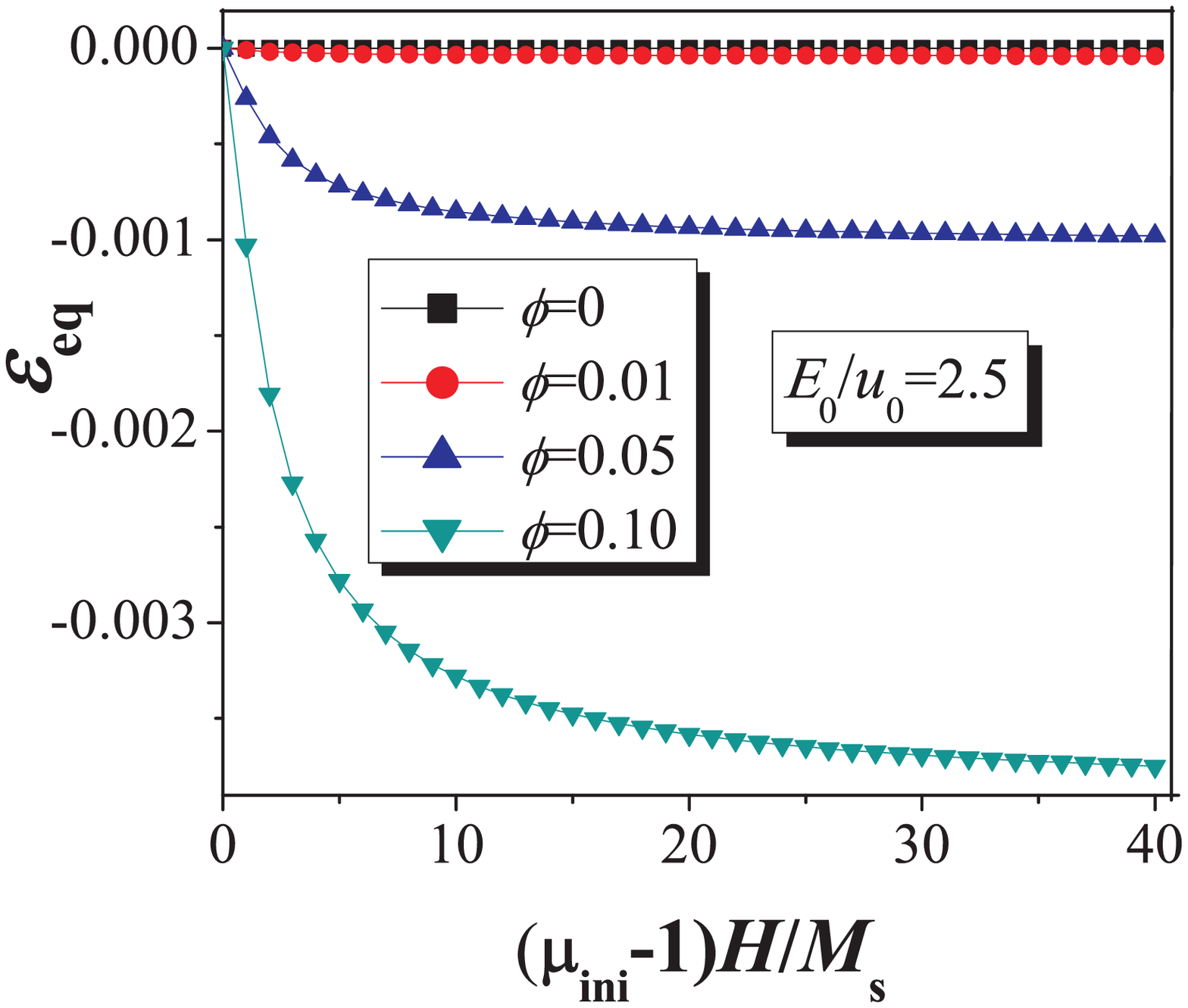}}
\hspace{0.17\textwidth} (a) \hspace{0.31\textwidth} (b) \hspace{0.3\textwidth} (c)
\caption{(Color online) Dependence of the equilibrium elongation
$\varepsilon_{\rm eq}$ on the reduced magnetic field
$h=(\mu_\mathrm{ini}-1)H/M_\mathrm{s}$ at different volume fractions $\phi$,
calculated for three types of initial lattices: simple cubic
lattice (a), body-centered cubic lattice (b) and hexagonal
close-packed lattice (c).}\label{fig5}
\end{figure}

\begin{figure}[!b]
\centerline{
\includegraphics[height=4.18cm]{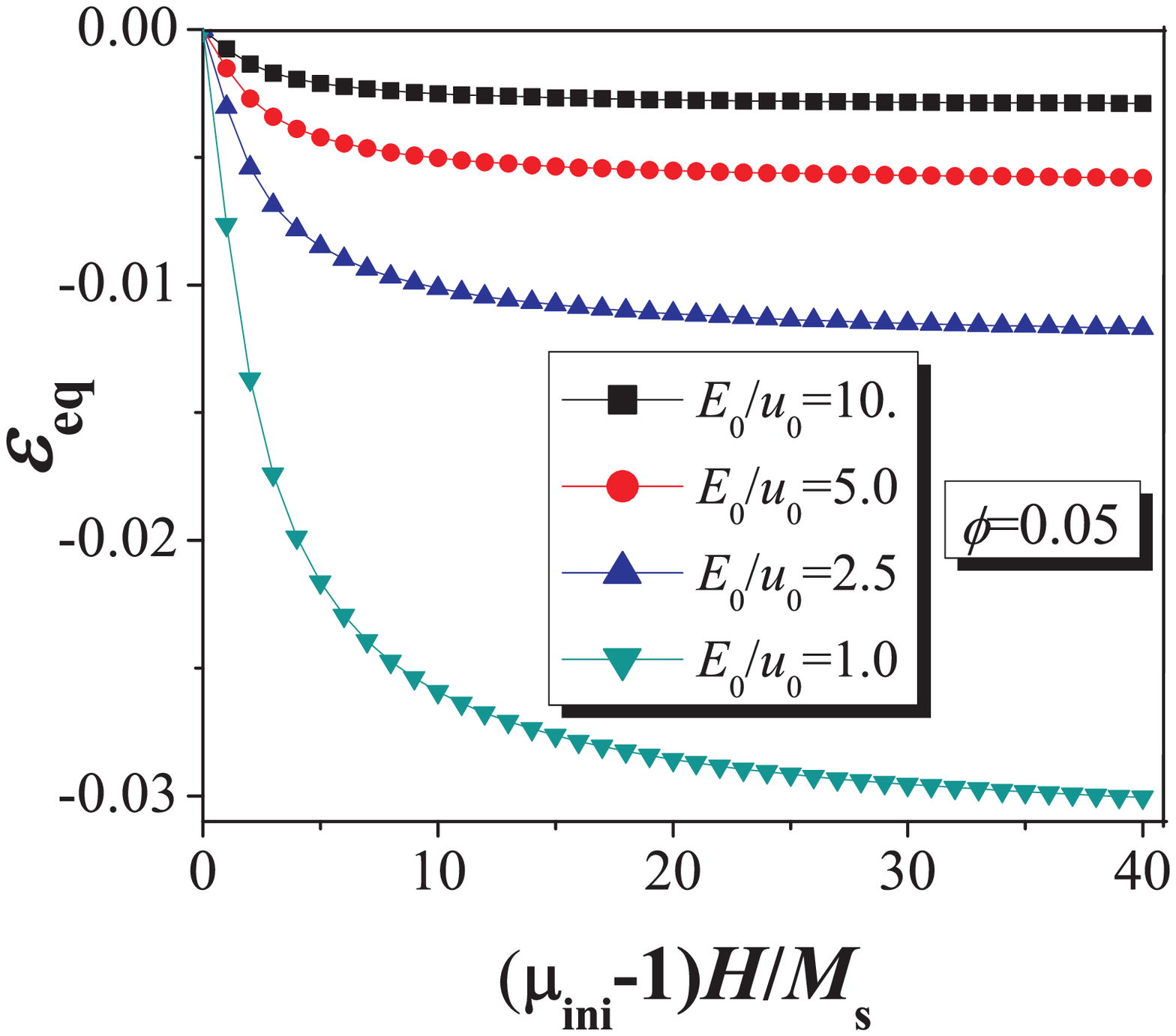}
\includegraphics[height=4.1cm]{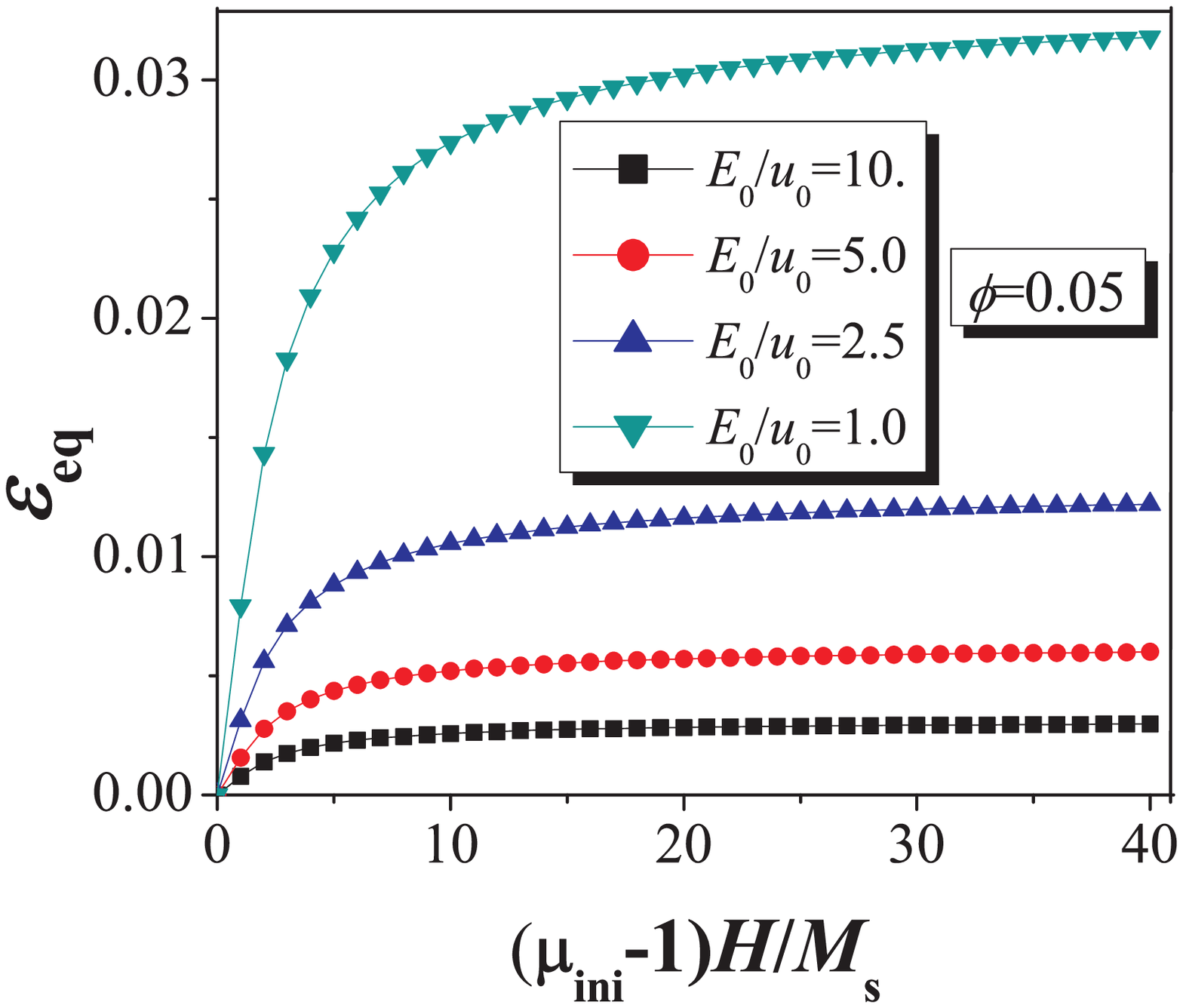}
\includegraphics[height=4.18cm]{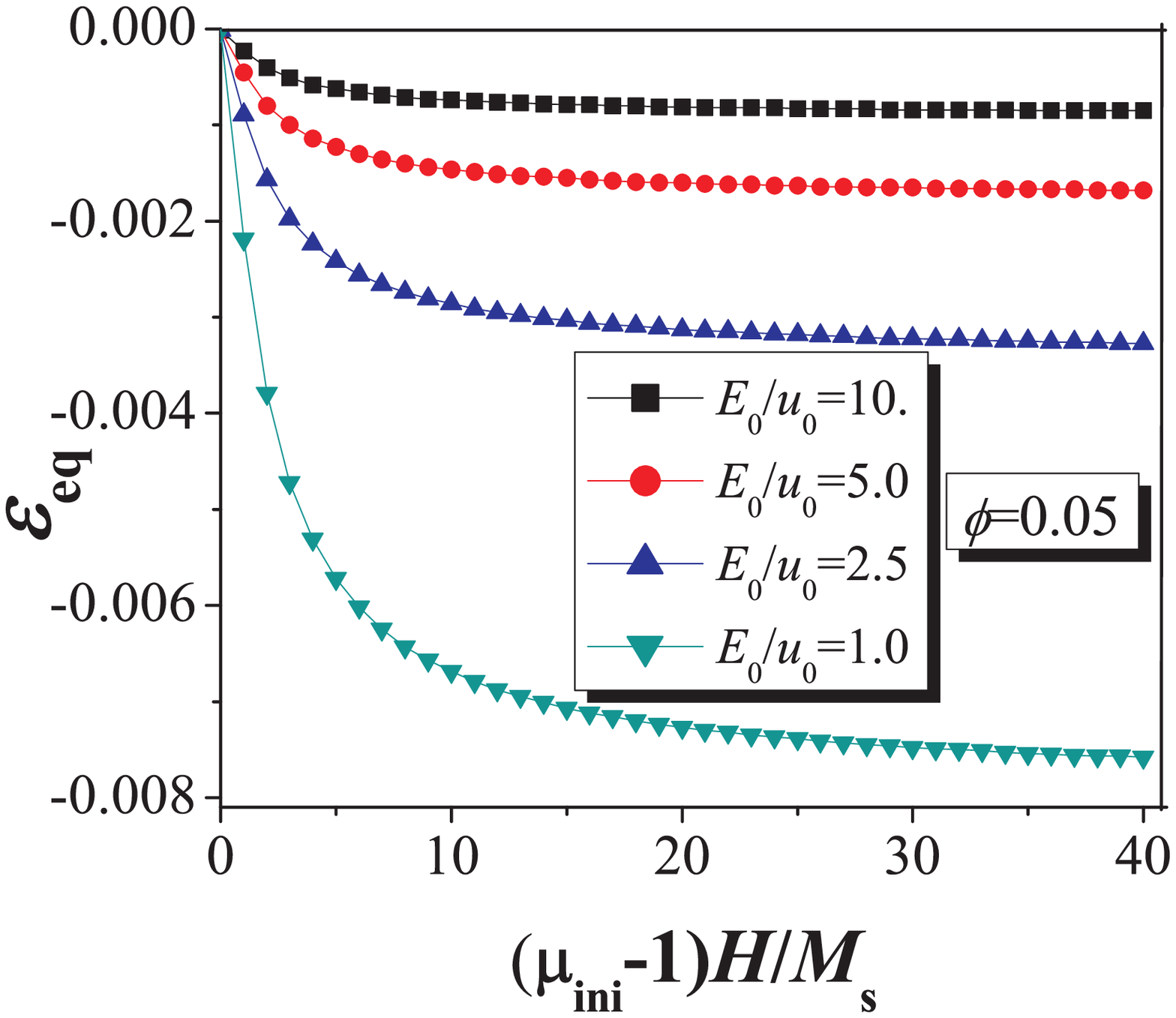}}
\hspace{0.17\textwidth} (a) \hspace{0.31\textwidth} (b) \hspace{0.3\textwidth} (c)
\caption{(Color online) Same as figure~\ref{fig5} but at different values of the
parameter $E_0/u_0$ and at fixed volume fraction
$\phi=0.05$.}\label{fig6}
\end{figure}

In this section we study the magnetostriction effect in the MSE
with isotropic spatial distribution of magnetic particles. We
calculate the equilibrium elongation of the MSE from the minimum
of free energy~$F$:
\begin{eqnarray}
\left. \frac{\partial F}{\partial
\varepsilon}\right|_{\varepsilon=\varepsilon_\mathrm{eq}}=0.\label{EFQ}
\end{eqnarray}
Using equations (\ref{FEN}) and (\ref{FFF}), the equation
(\ref{EFQ}) for the equilibrium elongation $\varepsilon_{\rm eq}$
can be rewritten as follows:
\begin{eqnarray}
\lefteqn{\frac{E_0}{3}\left[1+\varepsilon_{\rm
eq}-\frac{1}{(1+\varepsilon_{\rm eq})^2}\right]+u_0
\phi^2\left(\frac{h}{1+|h|}\right)^2\frac{1}{\gamma}\sqrt{1+\varepsilon_{\rm
eq}}}\nonumber\\&&\mbox{}\times\sum\limits_{r_i\neq
0}\frac{12(1+\varepsilon_{\rm
eq})^6(r_i)_x^4-30(r_i)_x^2(1+\varepsilon_{\rm eq})^3\left[(r_i)_y^2+
(r_i)_z^2\right]+3\left[ (r_i)_y^2+(r_i)_z^2\right]^2}{2\left[(1+\varepsilon_{\rm
eq})^3(r_i)_x^2+(r_i)_y^2+(r_i)_z^2\right]^{\frac{7}{2}}}=0.\label{EE}
\end{eqnarray}
Dividing both the left- and right-hand sides of equation
(\ref{EE}) by the factor $u_0$, one can see that the equilibrium
elongation $\varepsilon_{\rm eq}$ depends on the elastic modulus
$E_0$ and on the magnetic parameter $u_0$ through their
dimensionless ratio $E_0/u_0$. The last equation we solve
numerically at varied values of the reduced magnetic field $h$.
Figure~\ref{fig5} shows dependences of the equilibrium elongation
$\varepsilon_{\rm eq}$ on the reduced magnetic field  $h$ at the
values of parameter $E_0/u_0=2.5$ and volume fraction $\phi=0,
0.01, 0.05$ and $0.1$, calculated for the three types of lattice
models. Figure~\ref{fig6} shows dependences of the equilibrium
elongation $\varepsilon_{\rm eq}$ on the reduced magnetic field
$h$ at the values of volume fraction $\phi=0.05$ and parameters
$E_0/u_0=1.0, 2.5, 5$ and $10$, calculated for the three types of
lattice models. One can see that for simple cubic and
hexagonal close-packed lattices, an MSE is uniaxially contracted
along the direction of the external magnetic field,
$\varepsilon_\mathrm{eq}<0$ [figure~\ref{fig5}~(a), (c) and \ref{fig6}~(a), (c)],
while for the body-centered cubic lattice, an MSE uniaxially
expands along the direction of the external magnetic field,
$\varepsilon_\mathrm{eq}>0$ [figure~\ref{fig5}~(b) and \ref{fig6}~(b)].

\begin{wrapfigure}{i}{0.45\textwidth}
\centerline{
\includegraphics[width=0.4\textwidth]{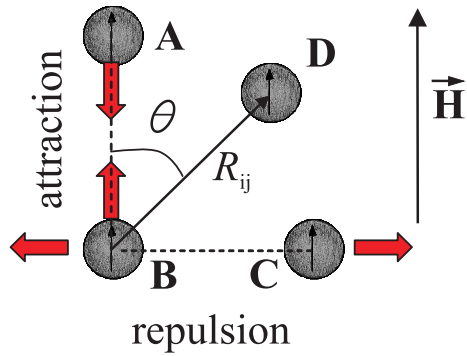}
}
\caption{\label{fig7} Attraction and repulsion of magnetic
particles in an MSE depending on their mutual positions.}
\vspace{-5mm}
\end{wrapfigure}

Different signs of magnetostriction of MSEs with different
spatial ``isotropic'' distributions of particles (simple cubic,
body-centered cubic and hexagonal close-packed lattices) can be
explained by the fact that the mechanical behaviour of MSEs is
determined mainly by the mutual attraction and repulsion of the
nearest particles as it is illustrated in figure~\ref{fig7}. In
a simple cubic lattice, the nearest particles to a given one (B)
are the particles A and C as presented in figure~\ref{fig7}. Such
configurations lead to the contraction of an MSE along the
magnetic field which is in accord with our results presented in figures~\ref{fig5}~(a)
and \ref{fig6}~(a). In the body-centered cubic and
hexagonal close-packed lattices there exist some extra particles whose
positions are determined by the angle $\theta$ (see figure~\ref{fig7}). Depending on the value of $\theta$, either contraction
or elongation of an MSE is possible. One can show that interaction
between the particles B and D results in the expansion at
$32^\circ<\theta<72^\circ$ and in contraction at
$0<\theta<32^\circ$, $72^\circ<\theta<90^\circ$. Therefore, the
body-centered cubic lattice predicts an expansion since $\theta_{\rm
BCC}=54.7^\circ$. For a hexagonal close-packed lattice, the
nearest particles
%
%
which lie at the angles $\theta_{\rm
HCP}=35^\circ$ provide an expansion and the nearest particles which
lie at $\theta_{\rm HCP}=90^\circ$ provide a contraction. We have
shown that the interplay between two different contributions of the
nearest particles leads to the contraction. Thus, the sign of
magnetostriction strongly depends on the local spatial
distribution of magnetic particles.

Furthermore, one can see in figure~\ref{fig5} that the increase
of the volume fraction $\phi$ results in an increase of the
equilibrium elongation $|\varepsilon_{\rm eq}|$, when $h$ is
fixed. Figure~\ref{fig6} shows that the increase of the parameter
$E_0/u_0$ results in a decrease of matrix deformation
$|\varepsilon_{\rm eq}|$, when $h$ is fixed. These results are
explained by the fact that the relative contribution of magnetic
interaction becomes larger at higher values of $\phi$ and at lower
values of $E_0/u_0$.

\subsection{Young's modulus of the MSE}

In the case of a tensile deformation, we consider such geometry
 when the mechanical force is applied along the
external magnetic field $\vec{\textbf{H}}$, i.e., along the
$x$-axis in our case (see figure~\ref{fig2}). The Young's modulus
$E$ can be obtained as the second derivative of free energy
with respect to $\varepsilon$, where
$\varepsilon=\varepsilon_\mathrm{eq}+\delta \varepsilon$:
\begin{equation}E =
\left.\frac{\partial^2 F}{\partial
\varepsilon^2}\right|_{\varepsilon=\varepsilon_\mathrm{eq}},\end{equation}
which yields
\begin{eqnarray}
\lefteqn{E=\frac{E_0}{3}\left[1+\frac{2}{(1+\varepsilon_{\rm
eq})^3}\right]-u_0
\phi^2\left(\frac{h}{1+|h|}\right)^2\frac{3}{4\sqrt{1+\varepsilon_{\rm
eq}}}
\sum\limits_{r_i\neq 0}
\Bigg\{{\left[32(r_i)_x^6(1+\varepsilon_{\rm
eq})^9
-\left[(r_i)_y^2+(r_i)_z^2\right]^3
\right.}\qquad\qquad}
\nonumber\\
&&
\left.
{}-192(r_i)_x^4\left[(r_i)_y^2\!+\!(r_i)_z^2\right](1\!+\!\varepsilon_{\rm
eq})^6
\!+\!90(r_i)_x^2\left[(r_i)_y^2\!+\!(r_i)_z^2\right]^2(1\!+\!\varepsilon_{\rm
eq})^3
\right]
%
{\left[(r_i)_x^2(1\!+\!\varepsilon_{\rm
eq})^{3}\!+\!(r_i)_y^2\!+\!(r_i)_z^2\right]^{-\frac{9}{2}}}\Bigg\}.\label{EEEE}
\nonumber\\
\end{eqnarray}
One can see that the ratio $E/E_0$ depends on the parameters $E_0$
and $u_0$ through their dimensionless ratio $E_0/u_0$.

\begin{figure}[!h]
\centerline{
\includegraphics[height=4.01cm]{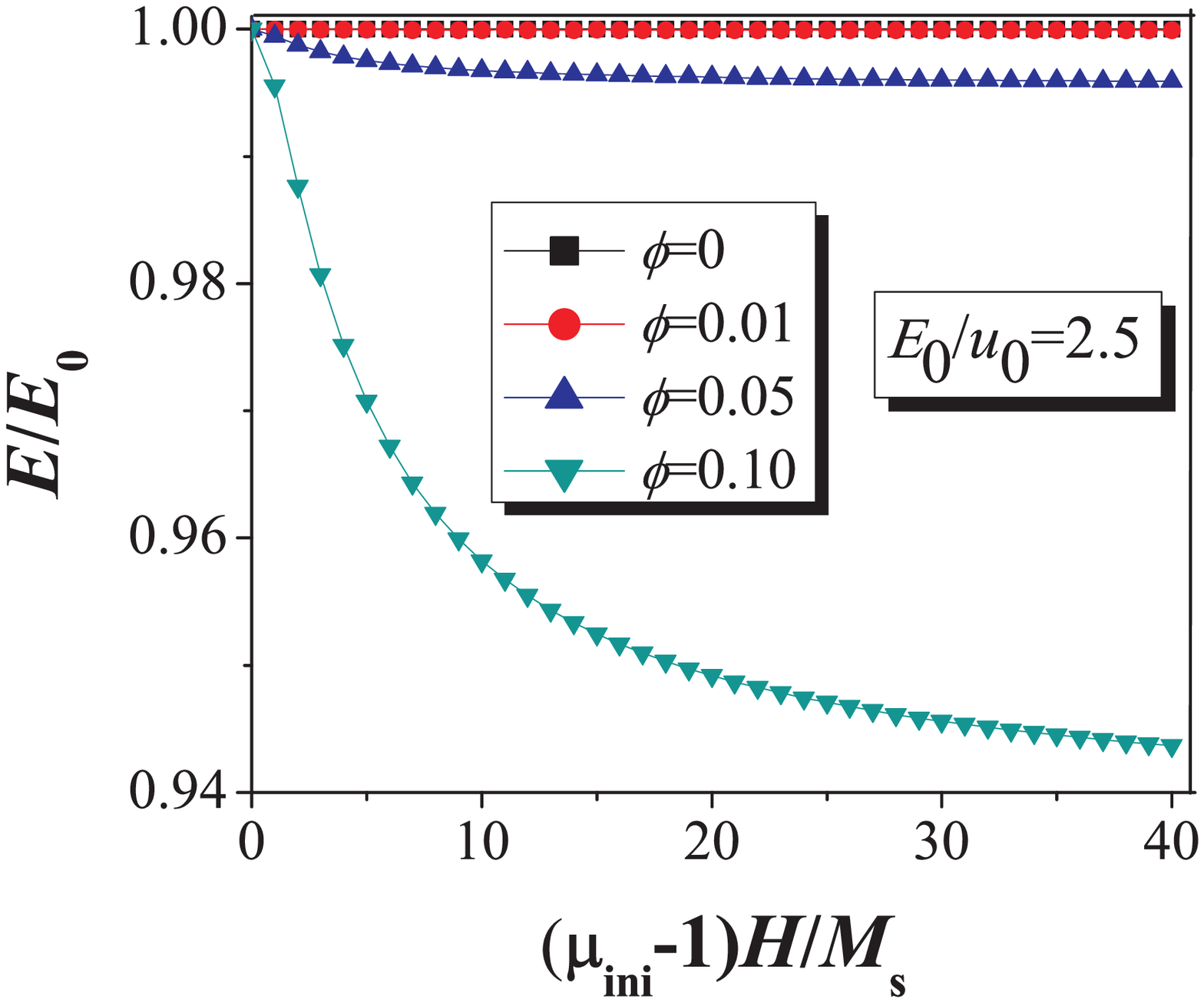}
\includegraphics[height=4.01cm]{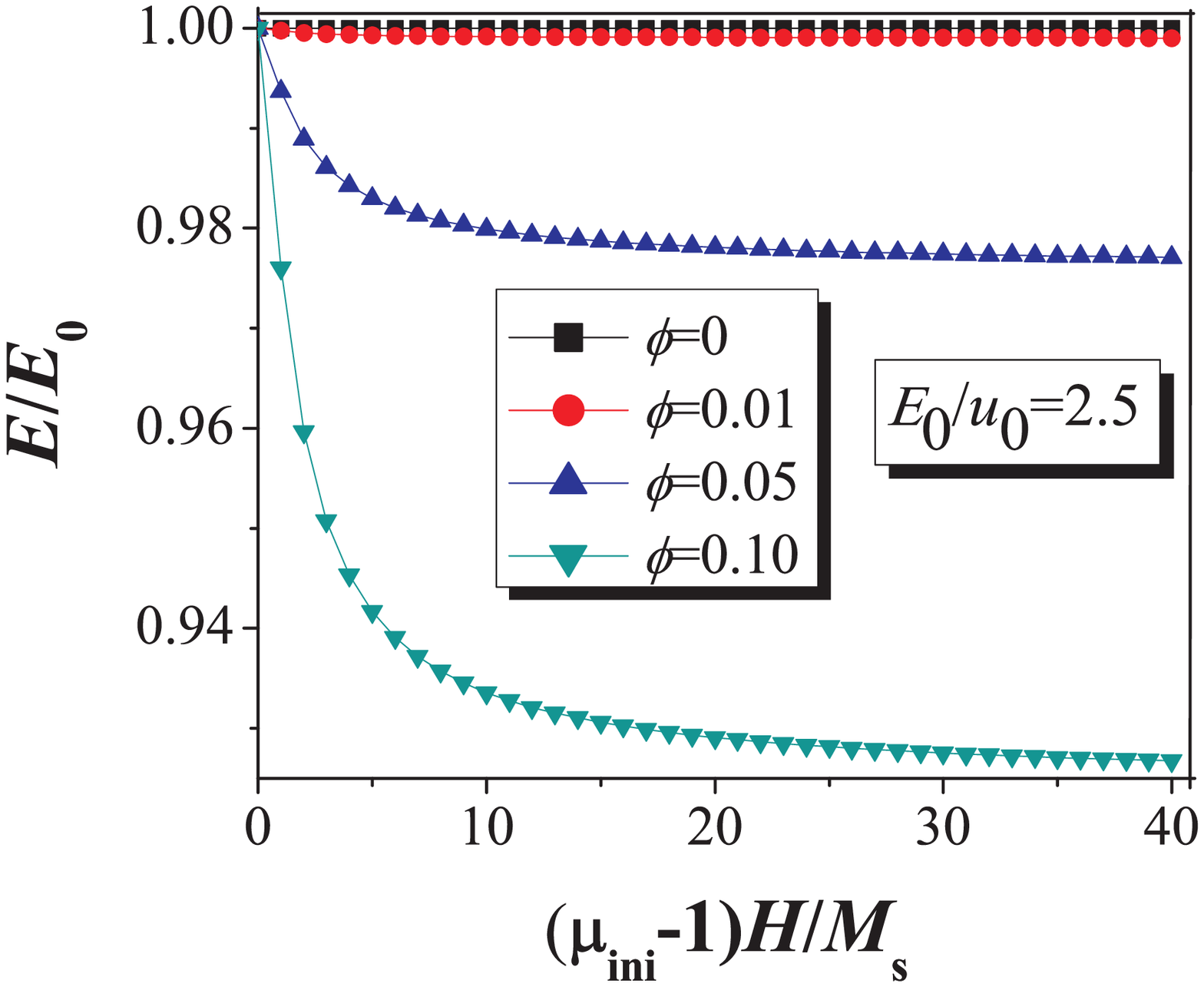}
\includegraphics[height=4.01cm]{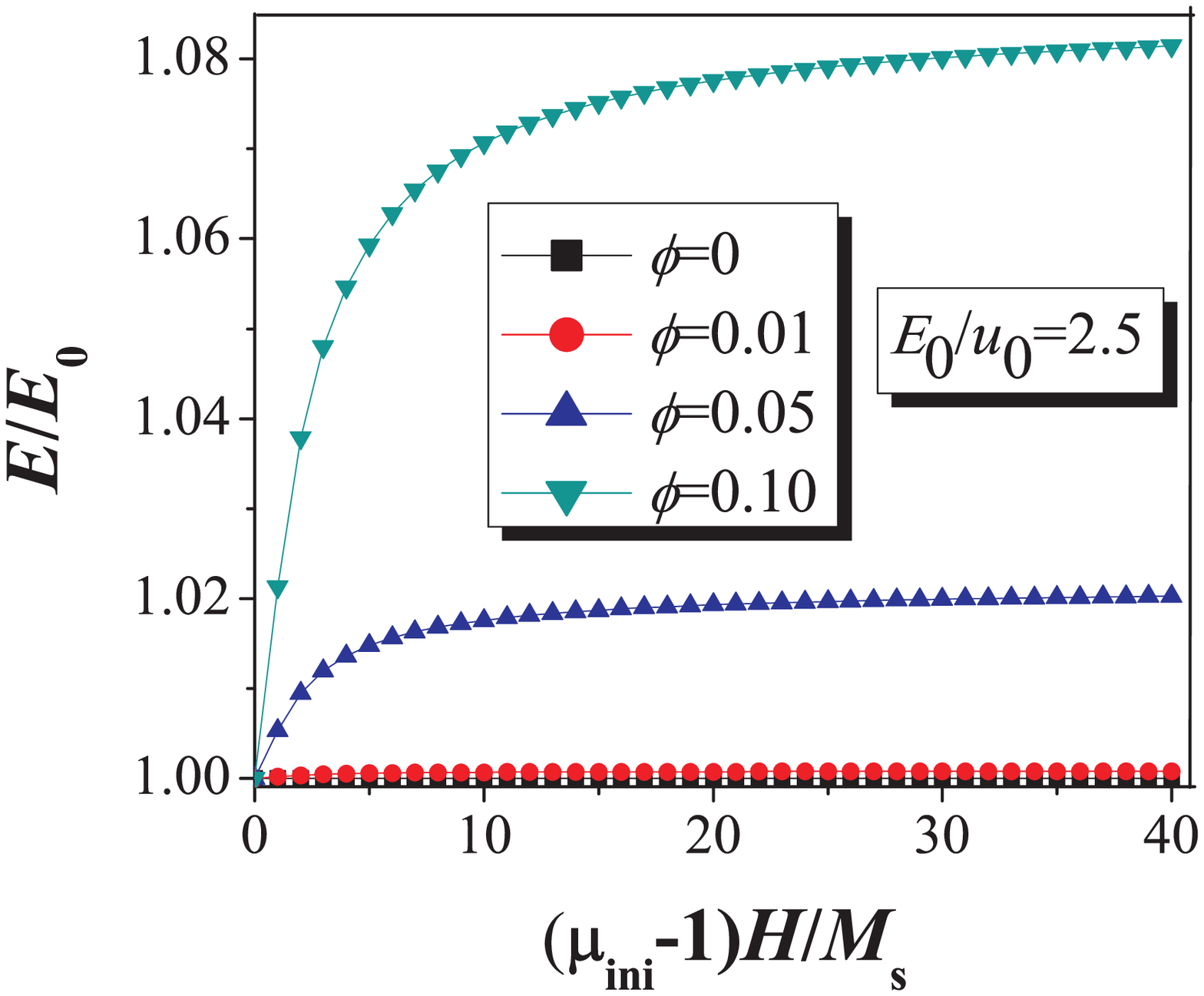}}
\hspace{0.17\textwidth} (a) \hspace{0.31\textwidth} (b) \hspace{0.3\textwidth} (c)\\
\caption{(Color online)
Dependence of the Young's modulus $E$ on the reduced magnetic
field $h=(\mu_\mathrm{ini}-1)H/M_\mathrm{s}$ at different volume fractions
$\phi$, calculated for three types of initial lattices: simple
cubic lattice (a), body-centered cubic lattice (b) and hexagonal
lattice (c).}\label{fig8}
\end{figure}

\begin{figure}[!t]
\vspace{3mm}
\centerline{\includegraphics[height=4.01cm]{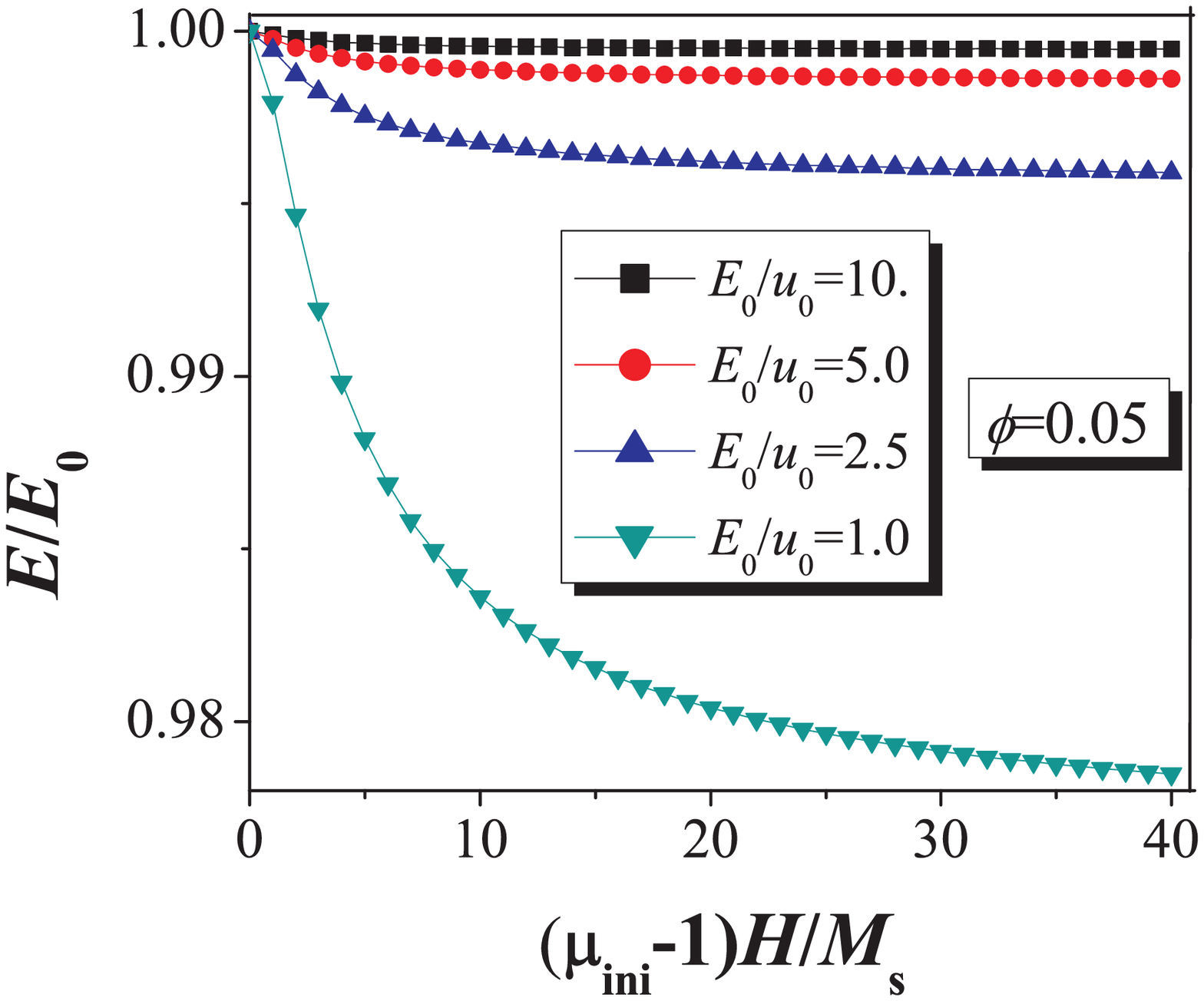}
\includegraphics[height=4.02cm]{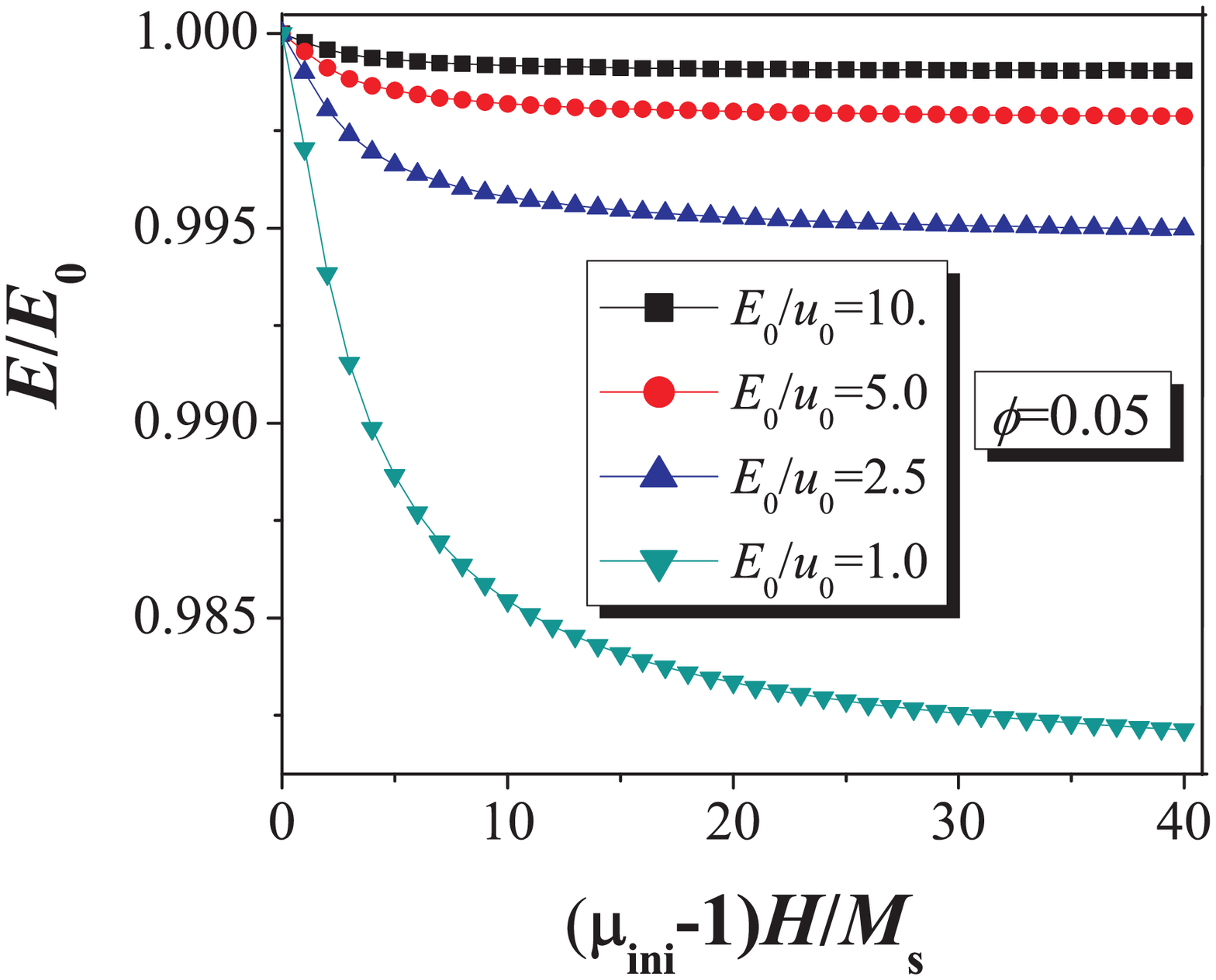}
\includegraphics[height=4.02cm]{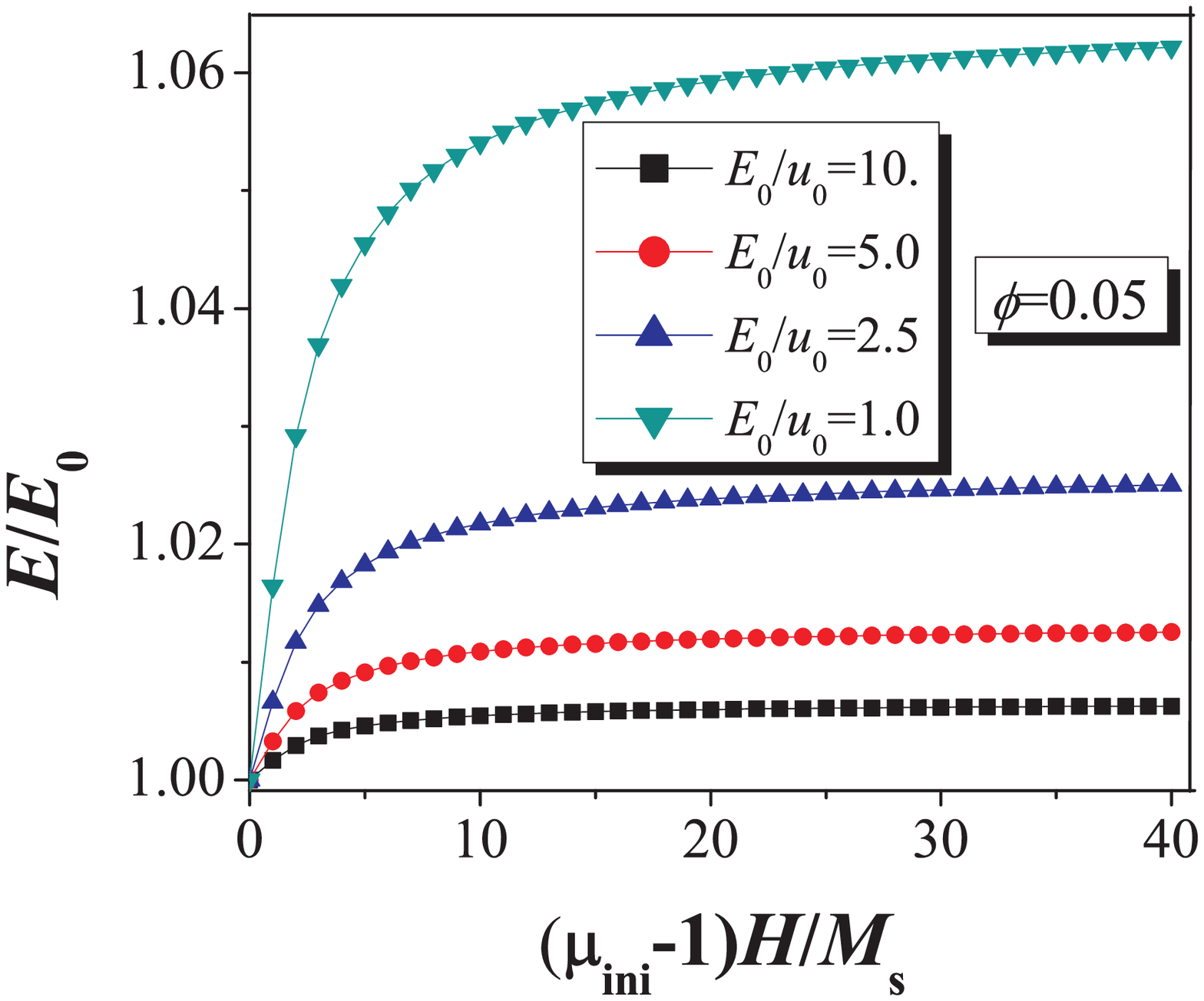}}
\hspace{0.17\textwidth} (a) \hspace{0.31\textwidth} (b) \hspace{0.3\textwidth} (c)
\caption{(Color online) Same as figure~\ref{fig8} but at different values of the
parameter $E_0/u_0$ and at fixed volume fraction
$\phi=0.05$.}\label{fig9}
\end{figure}

We have numerically calculated $E$ as a function of the reduced
magnetic field $h$ using equation~(\ref{EEEE}). Figure~\ref{fig8}
shows the dependence of the Young's modulus $E$ on the reduced
magnetic field $h$ at the values of parameter $E_0/u_0=2.5$ and
volume fraction $\phi=0, 0.01, 0.05$ and $0.1$, calculated for
three types of initial lattices: simple cubic, body-centered cubic
and hexagonal close-packed lattices.
 Figure~\ref{fig9} is the same
as figure~\ref{fig8} but at a fixed value of the volume fraction
$\phi=0.05$ and at varied values of $E_0/u_0=1.0, 2.5, 5$ and
$10$. One can see that the Young's modulus $E$ decreases for the
simple cubic [figure~\ref{fig8}~(a) and \ref{fig9}~(a)] and  the
body-centered cubic [figure \ref{fig8}~(b) and \ref{fig9}~(b)] lattices
and increases for the hexagonal close-packed lattice [figure~\ref{fig8}~(c)
and \ref{fig9}~(c)]. Different behaviour of $E$ as a
function of a reduced magnetic field $h$ for the three lattices
is caused by complex combinations of interactions between
particles of different mutual positions (see figure~\ref{fig7}).

Furthermore, it can be seen in figure~\ref{fig8} that an
increase of the volume fraction  $\phi$  leads to an increase of
absolute values of the change  of the modulus $|E-E_0|$ for
all distributions at fixed $h$. Figure~\ref{fig9} shows that an
increase of the parameter $E_0/u_0$ results in a decrease of
absolute values of the change  of the modulus $|E-E_0|$ for all
distributions at fixed $h$. These results are explained by the
fact that the relative contribution of the magnetic energy to the
modulus increases at increasing values of the volume fraction
$\phi$ and decreases at increasing values of the parameter $E_0/u_0$.

\vspace{3mm}
\section{Conclusion}

\looseness=1In this paper we have studied the mechanical properties of
magneto-sensitive elastomers with iso\-tro\-pic distribution of the
magnetic particles in an external magnetic field. We have used a
model in which magnetic particles are located at the sites of
regular lattices. Different types of the lattice models have been
considered: simple cubic, body-centered cubic and hexagonal
close-packed lattices. Such lattice distributions of particles can
be prepared in experiment as was shown in~\cite{Zhang08}. We show
that magneto-induced deformation and the Young's modulus of the
MSE strongly depend on spatial distribution of magnetic
particles. A simple cubic lattice model predicts a contraction of
the MSE with isotropic distribution of magnetic particles
along the direction of a homogeneous magnetic field. It predicts
that the Young's modulus decreases with an increase of the
magnetic field; the same result is obtained for the body-centered
cubic model. However, in contrast to the simple cubic lattice, the
body-centered cubic lattice model provides an expansion of the MSE
along the direction of a magnetic filed. The hexagonal lattice model
shows that MSEs with  isotropic particle distribution
contract along the external magnetic field, while its Young's
modulus increases. These findings may explain different signs of
magnetostriction observed in experiments with differently
prepared MSEs~\cite{Safronov11}.

\section*{Acknowledgements}

This work was supported by funds of European Union and the Free
State of Saxony (SAB project ECEMP B4 no. 13854/2379).

\ukrainianpart

\title{Вплив розподілу частинок на механічні властивості магнітночутливих еластомерів в однорідному магнітному полі}
\author{Д. Іванейко\refaddr{label1,label2}, В.П. Тощевіков\refaddr{label2,label3}, М. Саф'яннікова\refaddr{label2}, Г. Гайнріх\refaddr{label1,label2}}
\addresses{
\addr{label1} Інститут Матеріалознавства, Дрезденський Технічний
Університет, \\ вул. Гельмгольца, 7, 01069~Дрезден,  Німеччина
\addr{label2} Дрезденський Інститут полімерних досліджень ім.~Ляйбніца, вул. Гоге, 6,  01069~Дрезден, Німеччина \addr{label3}
Інститут високомолекулярних сполук, Російська Академія Наук, \\
Большой пр., 31, 199004 Санкт-Петербург, Росія}

\makeukrtitle

\begin{abstract}
\tolerance=3000%
Ми пропонуємо теорію, яка описує механічну поведінку
магнітночутливих еластомерів (МЧЕ) з ізо\-троп\-ним розподілом
частинок в однорідному магнітному полі. Використовуються три
механічні моделі, в яких магнітні частинки поміщено у вузлах
відповідно трьох періодичних ґраток: простої кубічної,
об'ємоцентрованої кубічної та щільно впакованої гексаґональної.
Цим ми розвиваємо наш попередній підхід [Ivaneyko~D. et al.,
Macromolecular Theory and Simulations, 2011, \textbf{20},
411], в якому ми використали лише просту кубічну ґратку для
опису просторового розподілу частинок. Деформація, яка виникає під
впливом магнітного поля, та модуль Юнга в МЧЕ розраховані для
різних напруженостей магнітного поля. Ми показали, що
магнітномеханічна поведінка МЧЕ сильно залежить від просторового
розміщення маг\-ніт\-них частинок. МЧЕ демонструє одновісне видовження
або стиск вздовж магнітного поля, а модуль Юнга зростає або спадає
при збільшенні напруженості магнітного поля залежно від
просторового розподілу магнітних частинок.

\keywords магнітночутливі еластомери, механічні властивості,
модуль, моделювання

\end{abstract}

\end{document}